C2SMART
CONNECTED CITIES WITH SMART TRANSPORTATION

WHITE PAPER ISSUE 2

# The Effects of the COVID-19 Pandemic on Transportation Systems in New York City and Seattle, USA

Jingqin Gao, Jingxing Wang, Zilin Bian, Suzana Duran Bernardes, Yanyan Chen, Abhinav Bhattacharyya, Siva Soorya Muruga Thambiran, Kaan Ozbay, Shri Iyer, Xuegang Jeff Ban

Contact: c2smart@nyu.edu
c2smart.engineering.nyu.edu

## Executive Summary

This paper, following up the white paper released on April 3, 2020, continues to highlight trends in mobility and sociability in New York City (NYC), and supplements them with similar data from Seattle, WA, two of the cities most affected by COVID-19 in the U.S. Seattle may be further along in its recovery from the pandemic and ensuing lockdown than NYC, and may offer some insights into how travel patterns change. Finally, some preliminary findings from cities in China are discussed, two months following the lifting of their lockdowns, to offer a glimpse further into the future of recovery.

*Key findings from NYC*

- NYC continues to see a consistent low volume in transit ridership and motor vehicle trips. Subway ridership remains down 91% and vehicular traffic via MTA bridges and tunnels was down 68% in April 2020 vs. 2019.
- An increase in traffic speed was observed with some city streets operating at or near free-flow speed. A 108% average increase in speeds on Midtown Avenues between 8AM-6PM was observed in April as compared with February. Concurrently, a 72% average daily increase in the number of school zone speeding tickets was observed during before and after stay-at-home orders.
- Crashes remain low as a result of the low vehicular volumes, but the severity of crashes measured by fatalities is up. This may possibly be due to higher vehicular speeds, but more data is needed.
- Weigh-in-motion (WIM) from the Brooklyn-Queens Expressway (BQE) showed reduced volumes and higher speeds as well as that the number of very heavy trucks (GVW > 100 kips) was reduced 30% for Queens bound and 44% for Staten Island bound traffic.
- Bikeshare ridership remains down with the exception of the first 12 days of March. Ridership patterns did change, with 15% fewer Friday and Saturday trips. Average trip duration in March 2020 also increased to 13.9 minutes, a 20% increase compared to the same month last year.
- An increase in social distancing complaints has made the *Non-emergency police matter* category the 2nd most common complaint among NYC311 reports.

*Key findings from Seattle*

- Public transit demand is still at a comparatively low level. Transit demand in Seattle continuously decreased until March 26 with a 79% drop compared to a typical weekday.
- Traffic volumes have begun to recover from previous lows. Data collected from downtown Seattle on I-5 show a 6.6%, 11.9%, and 14.8% increase in the weeks of April 6, 13, and 20, respectively, compared with March 30-April 5.
- Data showing that traffic volumes are starting to increase while public transit demand remains low indicates a potential shift in mode choice and increased preference for non-public modes of travel, at least in the near future.
- Bicycle volume counts near Fremont Bridge grew to a relatively high level (around 2,100) in April compared to March 2020, however, fewer day-of-week variations were observed compared to January and February 2020. Seattle opened 9 miles of Stay Healthy Streets from its bike facilities and trails to allow for safe social distancing, which may have contributed to the increase of bike usage in April.

## New York City

### Corridor Travel Time

The closing of essential businesses and stay-at-home policies have had a significant impact on average travel times in both NYC and Seattle from early March through the end of April. Travel time on Route 495 that connects Long Island and New Jersey via Queens and Manhattan in NYC was analyzed in Figure 1. A flatter travel time pattern of near free-flow speeds was observed through the data following the stay-at-home orders, instead of typical spikes of commuter peaks (Figure 1(b)). Most recently, there is some evidence of increasing travel times. For example, Westbound average travel time for the week of April 20th slightly increased to 30.0 minutes, compared to 27.6 minutes for the week of April 6th (the second week after the stay-at-home order), an increase of 8.5%. No significant increase was overserved for the eastbound traffic, therefore it is not yet clear if this data illustrates a trend.

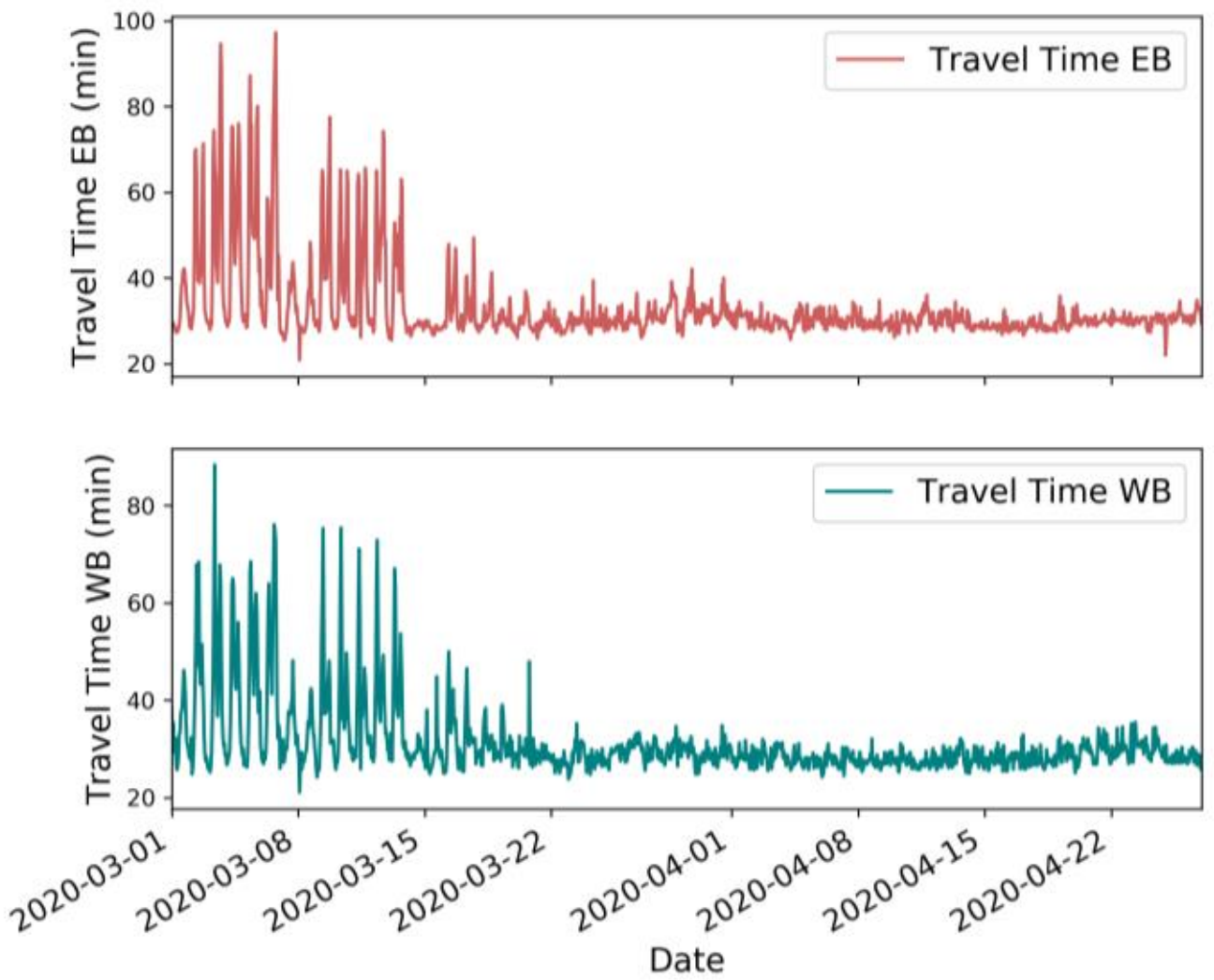

(a) Hourly travel time on the 495 corridor (March-April)

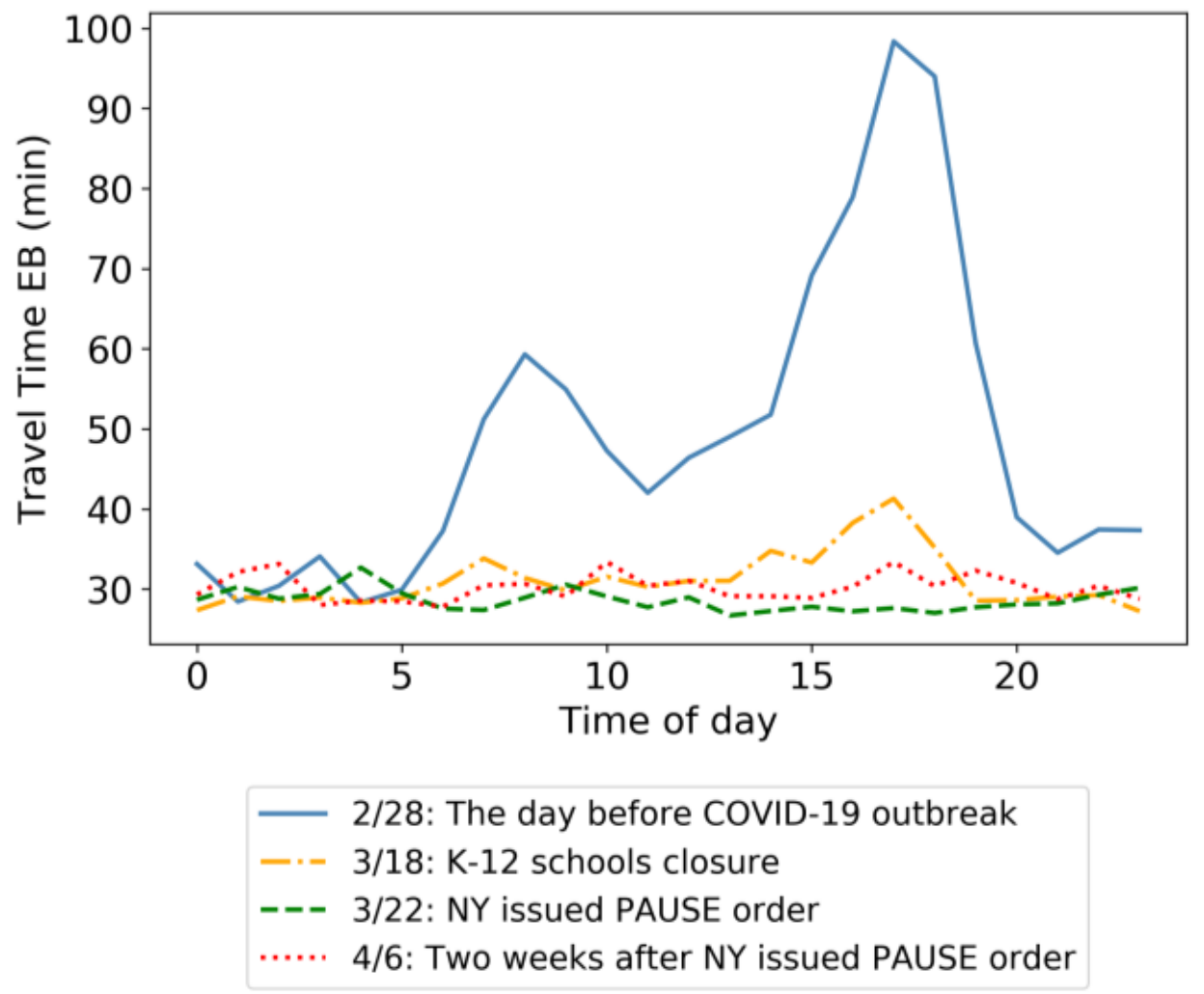

(b) Average hourly travel time on the 495 corridor

**Figure 1** Hourly travel time on 495 corridor (Flushing Meadows Corona Park to NJ Turnpike Exit 16E)

## Citywide Traffic Speeds

The speed maps in Figure 2 display the median travel speeds on major arterials/expressways of NYC as well as the Midtown Manhattan street network. With less traffic in the city, most roads and highways had higher speeds in the third week of April, compared to the same week in February. For example, on April 15, during the traditional morning peak hour, 8.3% of the 145 local road segments in Manhattan where data is available, had an average traffic speed over the city speed limit of 25 mph. 28% of them were over 20mph, while only 3.4% had an average traffic speed over 20mph in February, before the pandemic.

Average traffic speeds from 8AM to 6PM on north-south avenues between 34th and 57th Streets in Midtown Manhattan are shown in Figure 3. The average traffic speeds on avenues were up 108% in the third week of April, with a range from 22% to 253% on different avenues, compared to average traffic speed in the third week of February 2020. The average traffic speed on major crosstown streets, including 14th, 23rd, 34th, 42nd, and 57th Street, saw a 2-6 mph increase in the third week of April 2020 compared to the same week in February 2020.

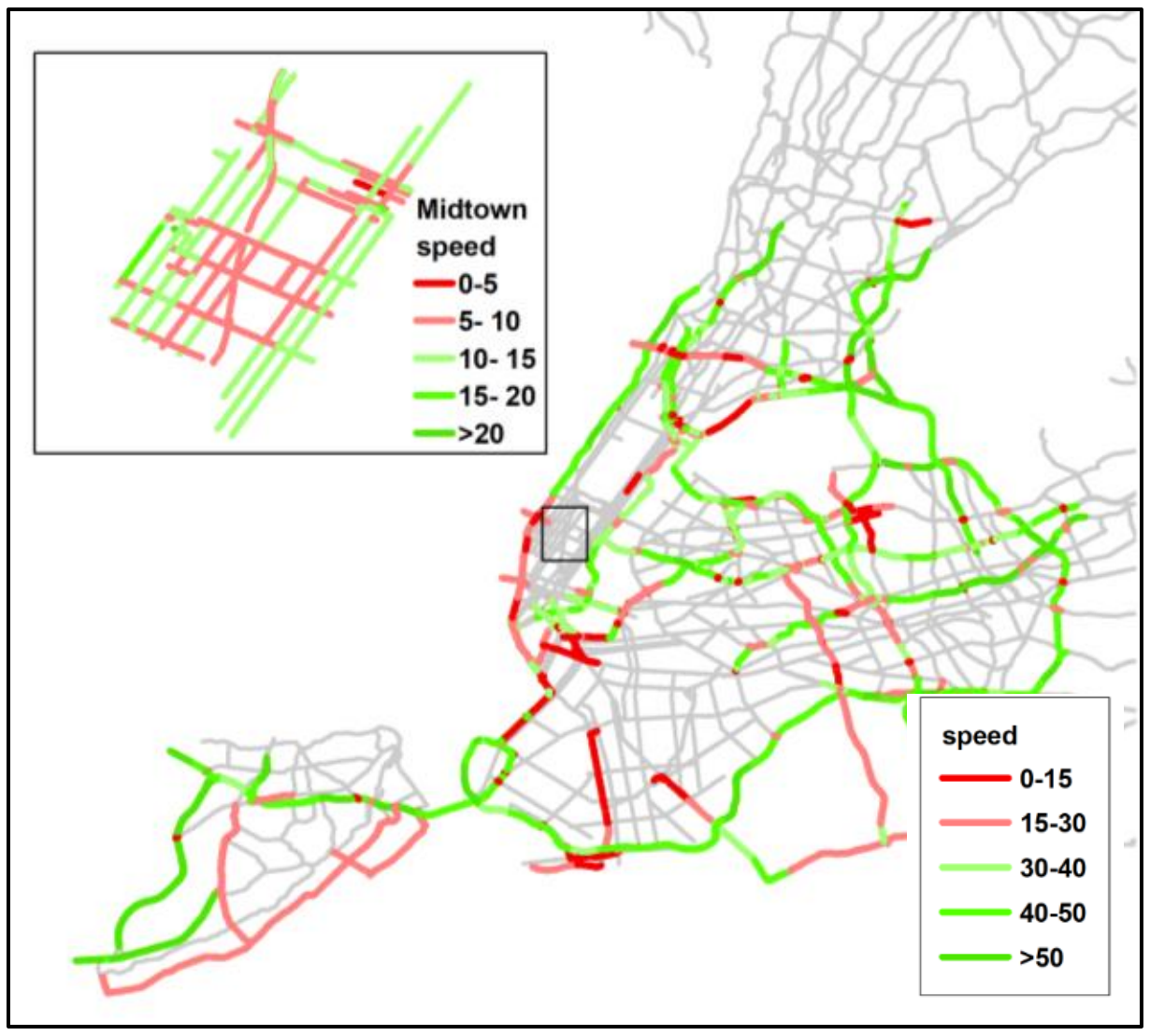

(a) Citywide average speeds February 17-21, 8AM to 6PM

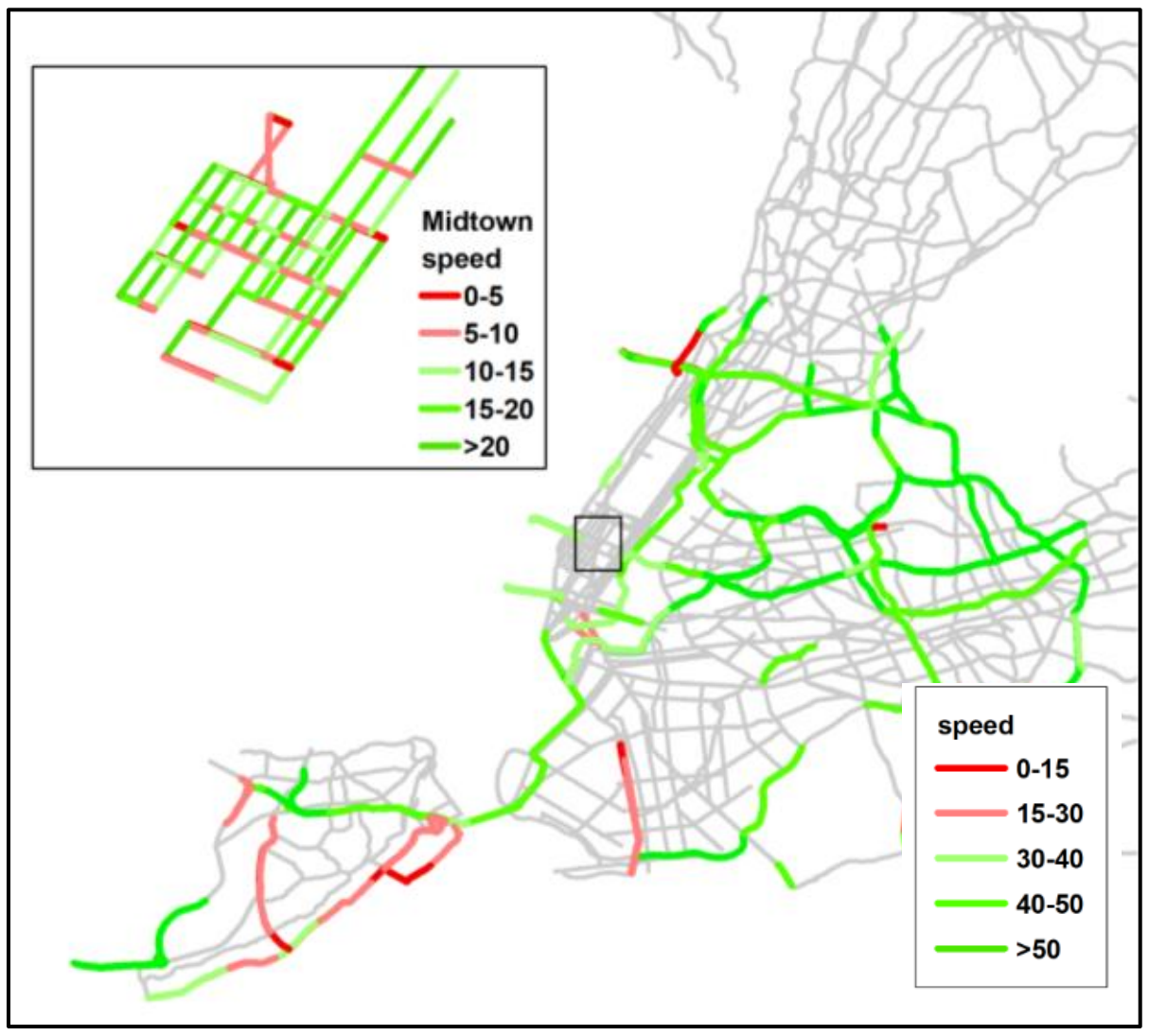

(b) Citywide average speeds April 20-24, 8AM to 6PM

**Figure 2** NYC Citywide average speeds, 8AM to 6PM

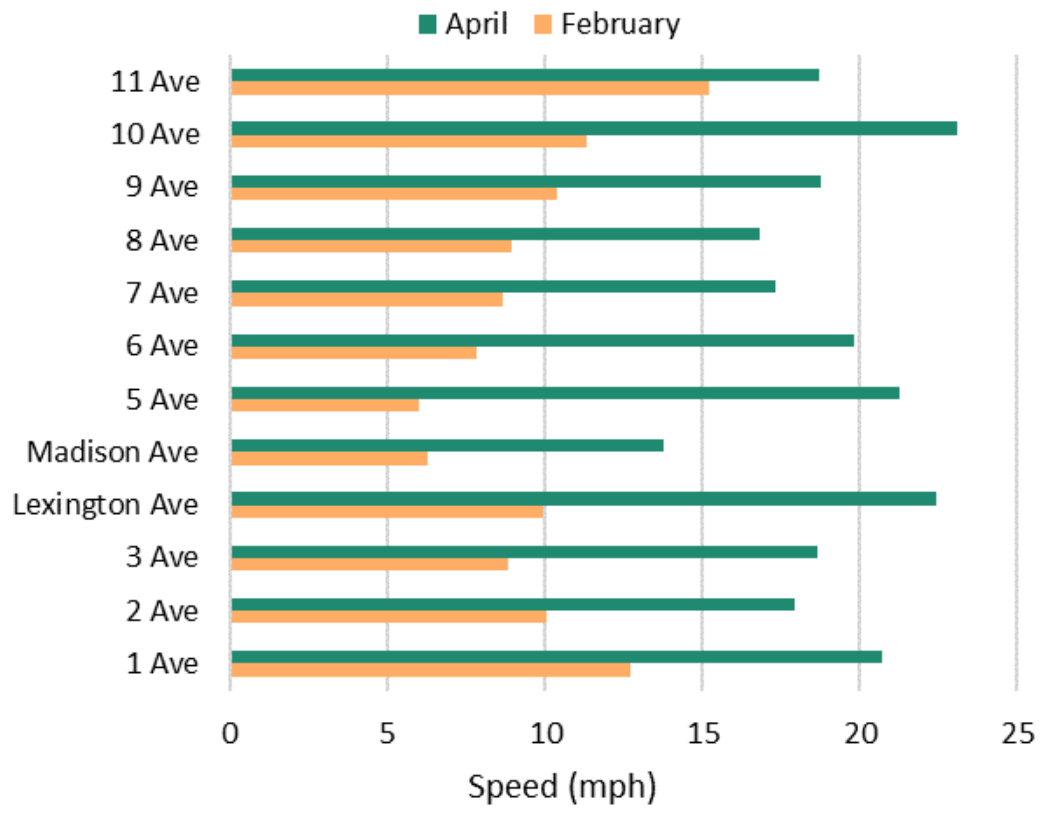

**Figure 3** Midtown Avenue speeds, 8AM to 6PM

## School Zone Speeding Tickets and Crash Data

As traffic speeds are up citywide, violations issued by the city's school zone cameras also increased by 72% during March 13 to April 19 as compared to January 1 to March 12 in 2020.

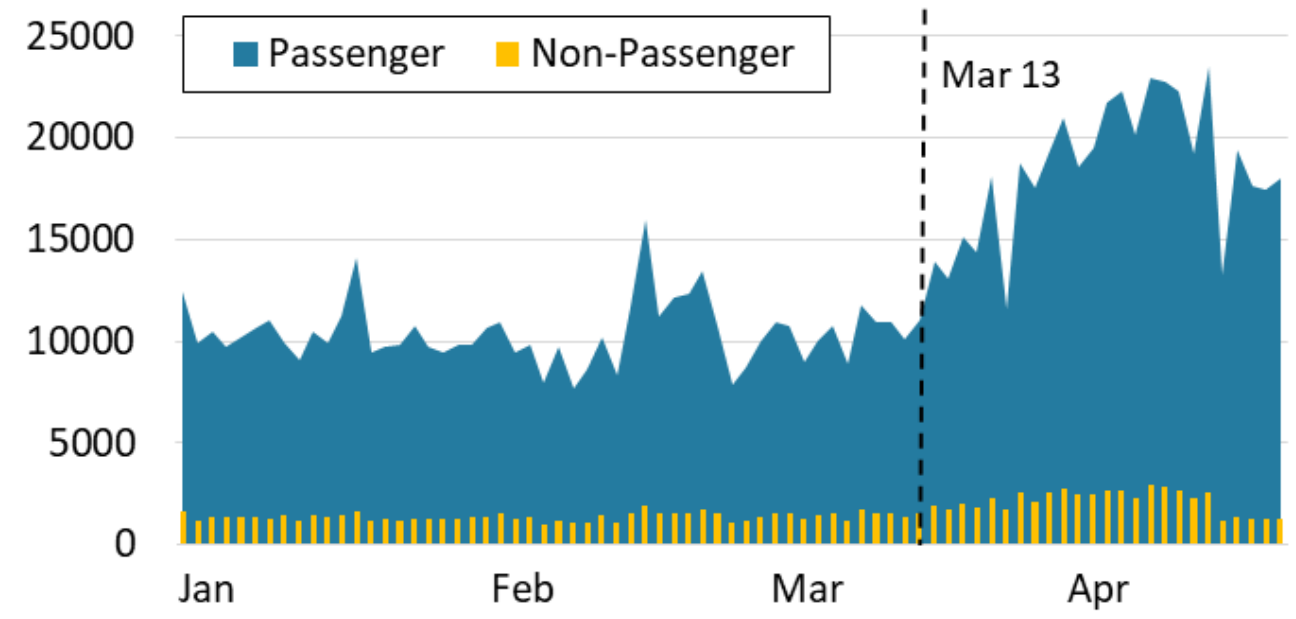

**Figure 4** Number of school zone speeding tickets from weekdays January 1 to April 19 in 2020.

Figure 4 shows the number of weekday school zone speeding tickets from January 1 to April 19 in 2020. There has been a sharp escalation in the number of the tickets, especially for passenger vehicles - a steep 75% increase after March 13 compared to the violations before March 13 - which indicates that drivers are taking advantage of fewer vehicles on streets to increase travel speeds. Figure 5 illustrates breakdown of school zone speeding tickets by borough. From March 23 to April 19, the number of tickets issued in Manhattan increased to almost four times the tickets issued in February 2020 (11,936 tickets in February and 46,691 tickets from March 23 to April 19).

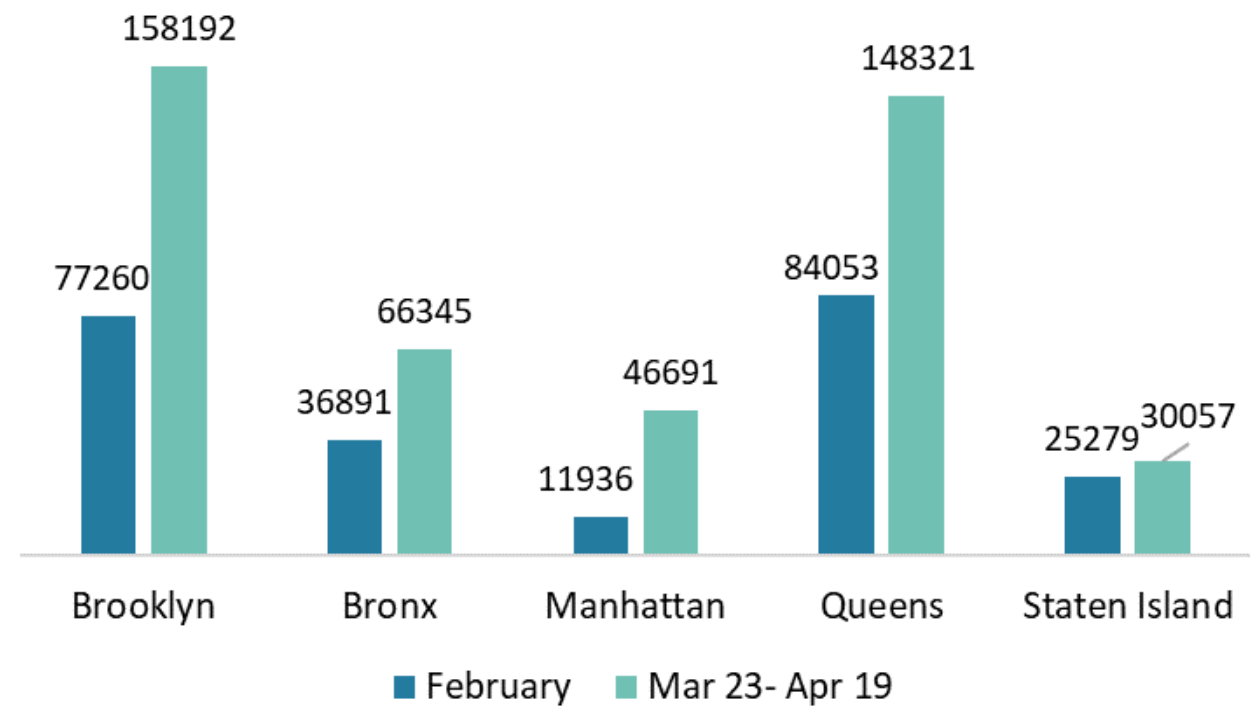

**Figure 5** School zone speeding tickets by borough

Some states have also seen an uptick in traffic fatalities[i]. Based on NYPD reported motor vehicle collisions, the number of reported crashes seen a continue decrease of 17%, 24%, and 8% in the first three weeks of April, 2020, compared to the same weeks in 2019 (Figure 6). Although there is a decrease in traffic-related fatalities due to lower traffic volumes, the fatality rate in crashes (fatalities/number of total crashes) increased from 1.4 fatalities/1000 crashes to 1.9 fatalities/1000 crashes in the first three weeks of April compared to the same period of February 2020, which could indicate crashes are more severe. Motorist and cyclist injuries in terms of the total injuries were up by 3% and 1% respectively, though the percentage of pedestrian injuries was down 4% in the first three weeks of April compared to the same period last year. The increase in speeding and crashes should be monitored closely.

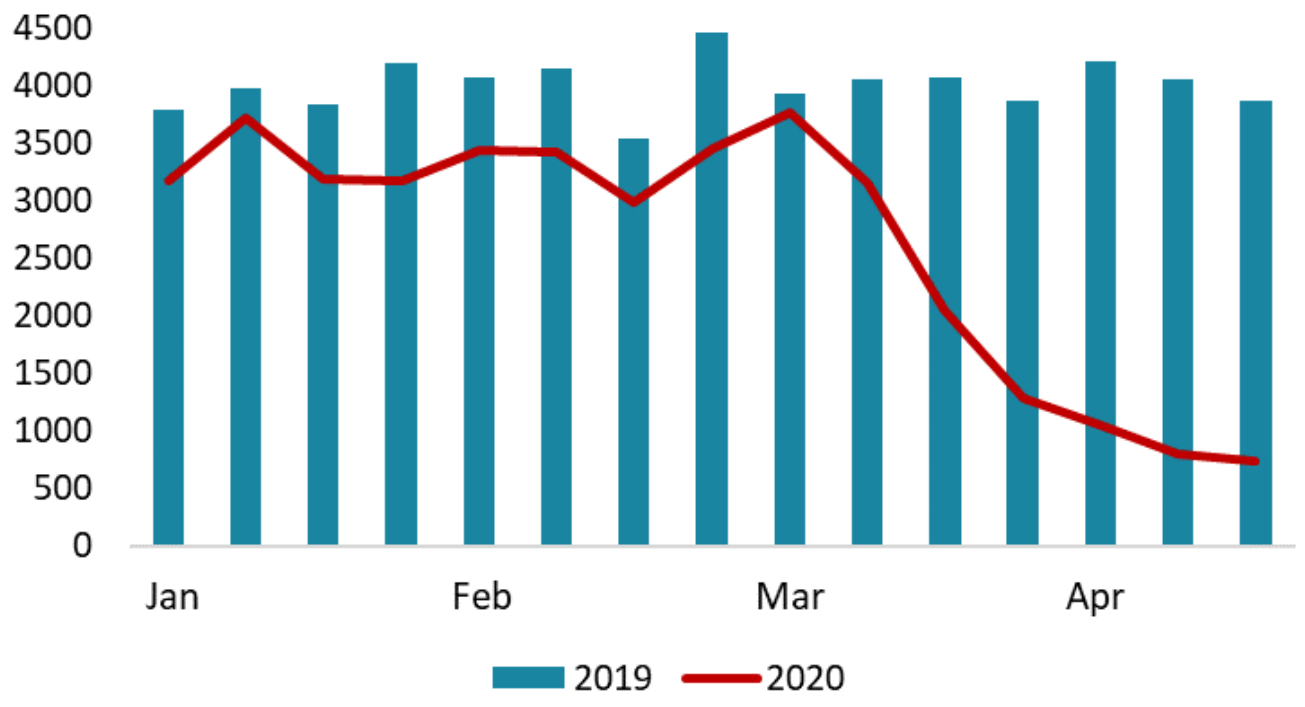

**Figure 6** Reported vehicle collisions in NYC by week

## Traffic Volume and Transit Ridership

In NYC, since the implementation of the stay-at-home order on March 22, both subway ridership and vehicular traffic on MTA facilities continues to be low. In the first three weeks of April, the subway ridership was down 91% and vehicular traffic via MTA bridges and tunnels was down 68%, compared to the same weeks in 2019. Recent data shows a smaller drop from 2019 levels, though there is yet to be an increase in absolute volumes.

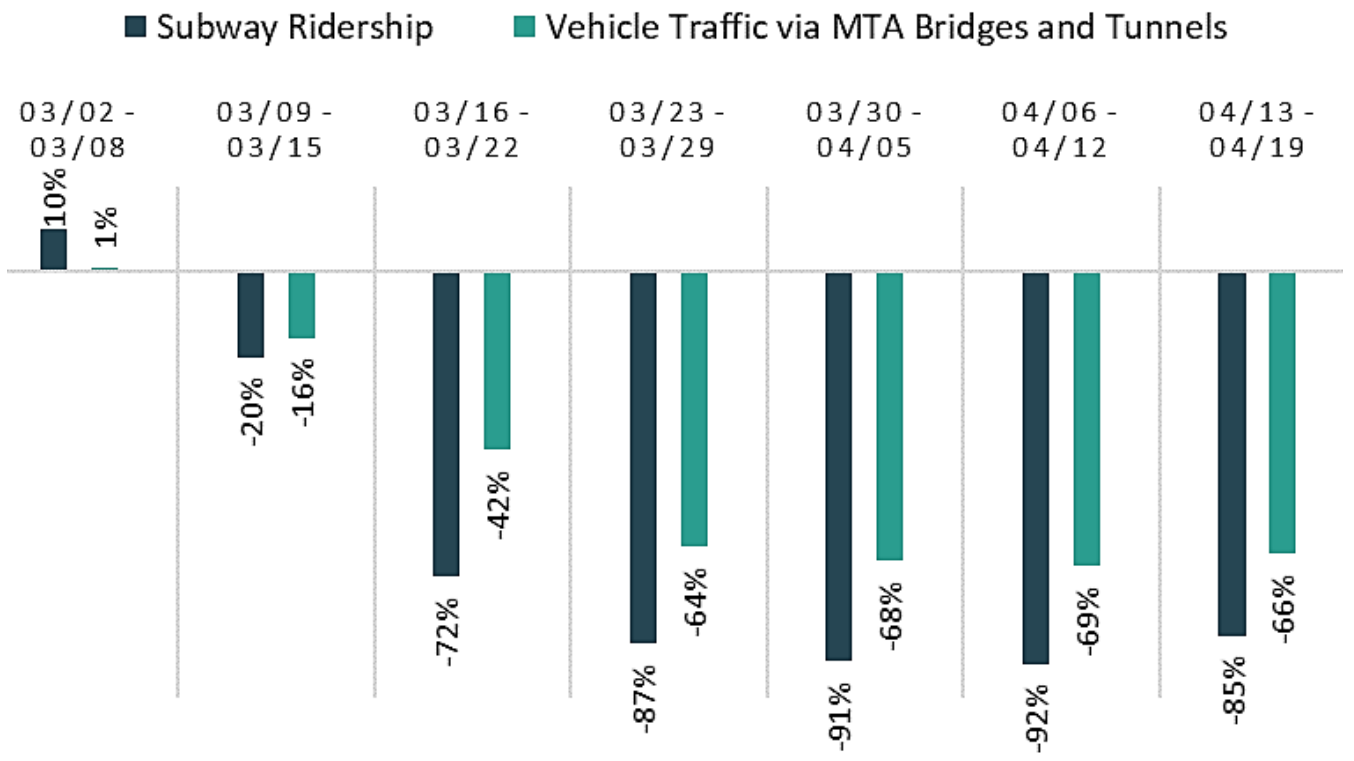

**Figure 7** Weekly changes in subway turnstile entries and traffic on MTA bridges and tunnels[ii]

## Weigh-in-Motion (WIM) Data from the BQE

WIM data from the urban roadway testbed on the Brooklyn-Queens Expressway (BQE) between Manhattan Bridge and Brooklyn Bridge show the following for February 3 to March 13 and March 13 to April 12, 2020:

- **Volume:** The annual daily traffic (ADT) for FHWA Class 1 through Class 3 was down by 39% and 37% for Queens bound (QB) and Staten Island bound (SIB), respectively, while the annual daily truck traffic (ADTT) was also down by 25% for QB and 30% for SIB.
- **Speed:** The average speed increased by 15% for QB (27 mph to 31 mph) and by 40% for SIB (20 mph to 28 mph).

- **GVW:** The average GVW for FHWA Class 6 and FHWA Class 9 was down by 3~8% for QB and SIB. The number of very heavy trucks (GVW > 100 kips) was down 30% for QB and 44% for SIB traffic.

### Citi Bike Ridership

Citi Bike is the largest bikeshare program in NYC. In the month of March, Citi Bike ridership increased more than 50% for the first 12 days of the month but then decreased significantly after March 16, when the "Stay at Home" order was issued, reaching a 62% drop in the total number of rides for the rest of the month when compared to 2019. The average ride duration in March 2019 was 11.6 minutes, while for March 2020, it was 13.9 minutes (+20%). After March 16, 2020, this increase was greater than 30%. The distribution of the duration of rides is also more widely dispersed in 2020, with a higher frequency of trips of more than 20 minutes. While the total number of rides tended to increase towards the end of the week in March 2019, it tended to decrease towards the end of the week in March 2020 (Figure 8). The fact that weekdays registered a higher total number of rides in March 2020 might indicate more rides being made for commuting or fewer recreational trips are being taken. These changes may indicate a mode choice change, with bicycle tris replacing transit, private vehicles, taxi, or TNC rides.

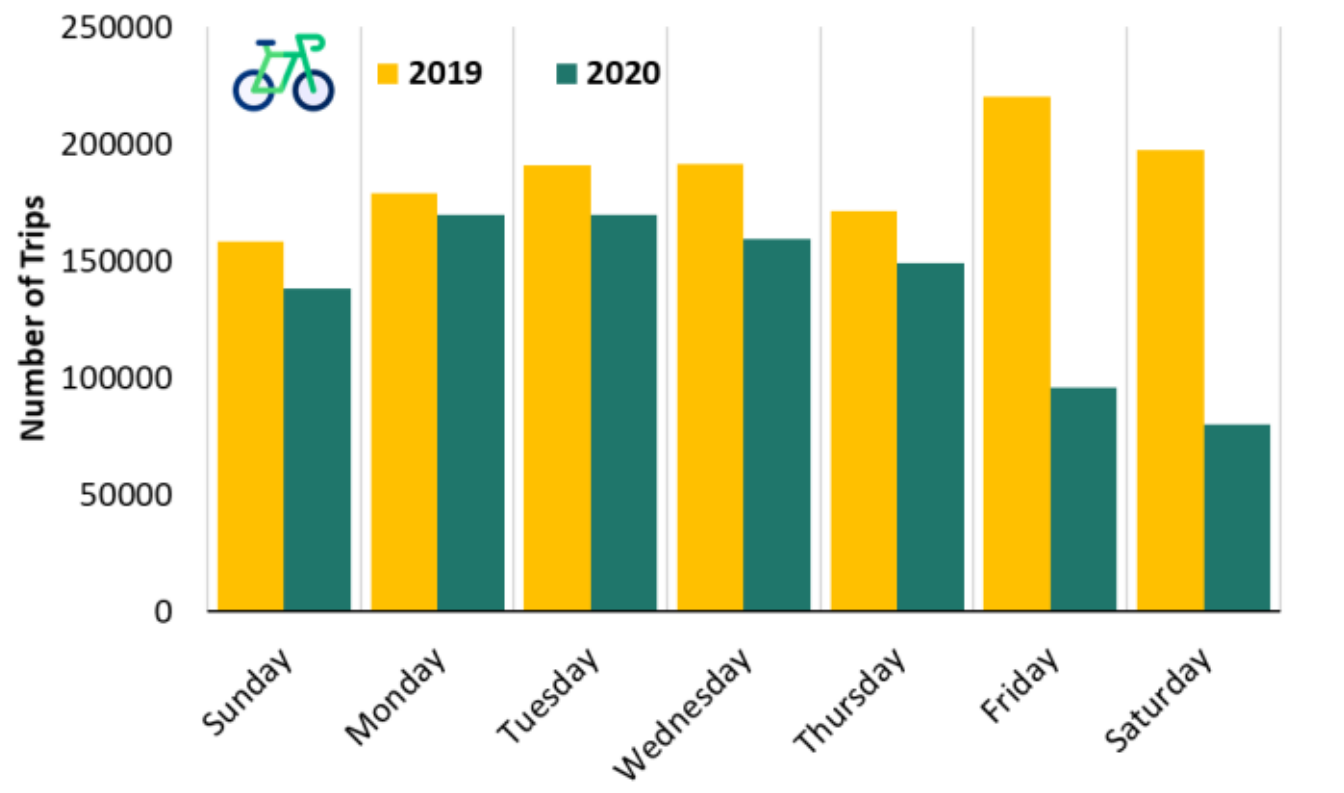

**Figure 8** Citi Bike trips per weekday

### NYC311 Service Requests

The 311 service line provides a way for NYC residents to find information about services, make complaints, and report problems. Not following social distancing guidelines currently carries a penalty of $1,000. On March 28, a new category called "Social Distancing" was added as a subtype under the "Non-Emergency Police Matter" complaint in NYC311. A complaint could be made for violation of social distancing rules to report a business or location that is required to be closed, an essential business that is open but is not complying with necessary restrictions or overcrowding at a business or location.

The number of reports to 311 surged starting from the first day of stay-at-home, with most of the complaints being directed at stores or commercial buildings, followed by residential buildings, streets/sidewalks, and playgrounds. A steady increase of complaints in the following weeks made the "non-emergency police matter" category shoot up to the second spot among all complaint types reported to 311 (Table 1). There were a total of 17,470 such complaints made in under a month (March 28 to April 22), with 16,925 of them being about social distancing, reported from all five boroughs of New York City.

**Table 1** The top 10 complaint types in NYC311 before and after stay-at-home order

| | **Feb 22 - Mar 21, 2020** | **Mar 21 - Apr 22, 2020** |
|---|---|---|
| **Rank** | **Complaint Type** | **Complaint Type** |
| 1 | Noise- Residential | Noise- Residential |
| 2 | Heat/Hot Water | Non-Emergency Police Matter |
| 3 | Illegal Parking | Heat/Hot Water |
| 4 | Blocked Driveway | Consumer Complaint |
| 5 | Street Condition | Illegal Parking |
| 6 | Street Light Condition | Noise- Street/Sidewalk |
| 7 | Noise- Street/Sidewalk | Street Light Condition |
| 8 | General construction | Blocked Driveway |
| 9 | Consumer Complaint | Noise |
| 10 | Noise | Noise- Vehicle |

## Seattle

Seattle was the first major U.S. city to experience the drastic effects of the COVID-19 crisis. However, health concerns have abated quicker as compared to NYC, and data suggests that Seattle may have begun returning to normal transportation conditions. Observing Seattle data may help to indicate what is to come in NYC.

### Corridor Travel Time

Travel time on Interstate-5 (I-5) that goes through downtown Seattle area were analyzed in Figure 9. Similar to NYC, Seattle observed a flatter travel time pattern of near free-flow speeds through the data, instead of typical spikes of commuter peaks.

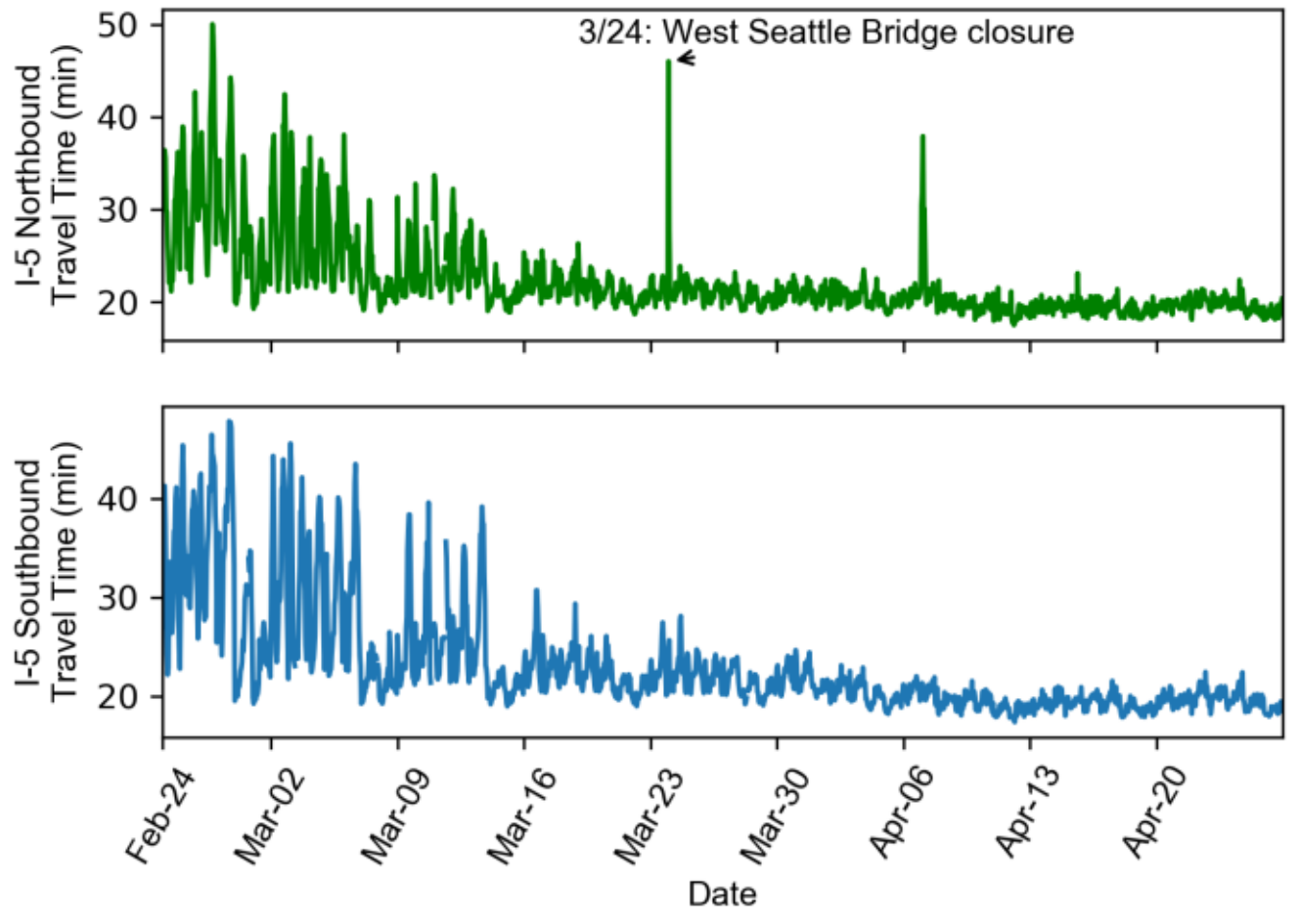

**(a)** Hourly travel time in March and April

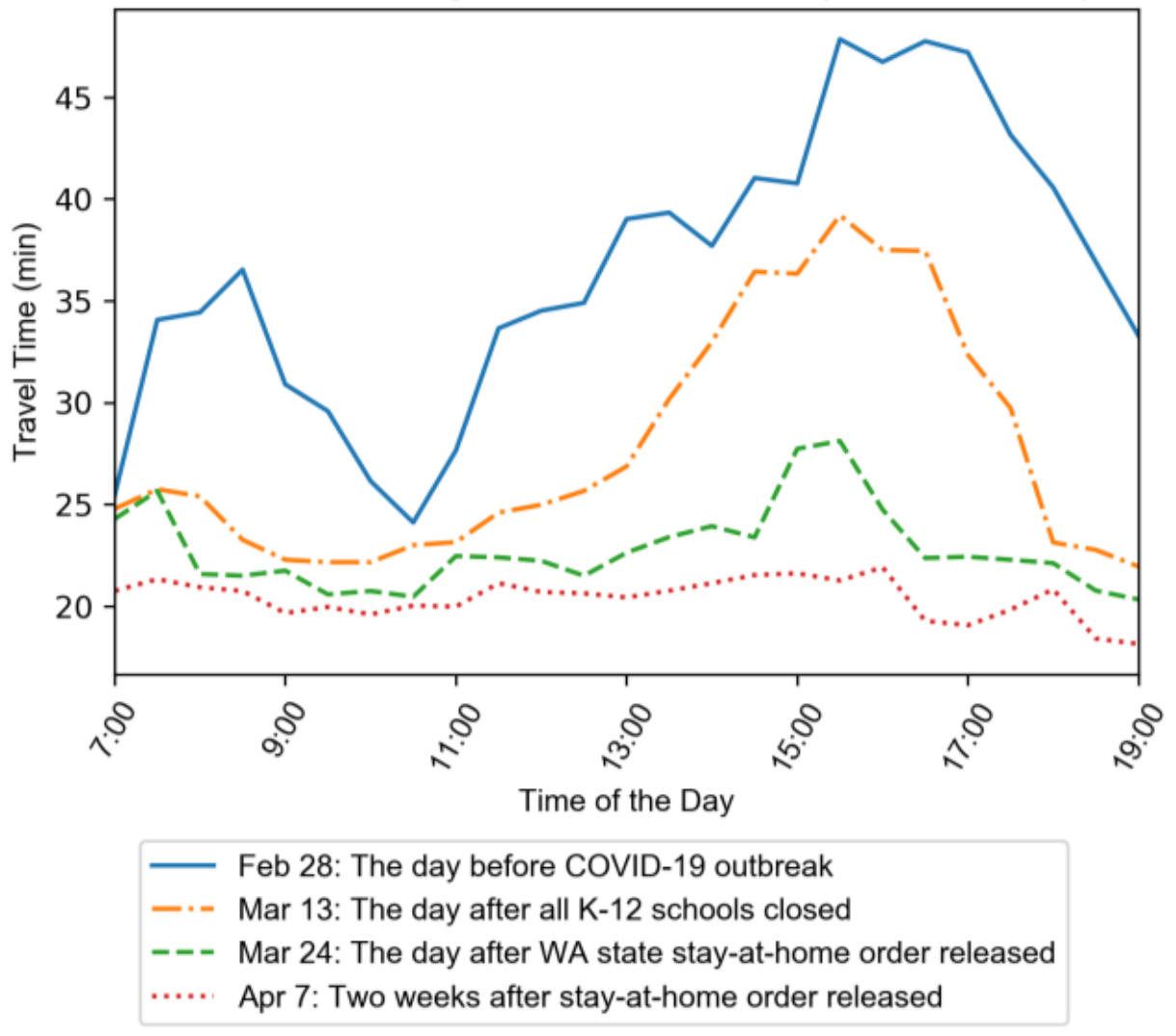

**(b)** Average hourly travel time on critical dates (I-5 SB)

**Figure 9** Hourly travel time on I-5 in Seattle

## Traffic Volume

Traffic volumes at three freeways locations (I-5 downtown Seattle, I-5 NE of Green Lake Park, and SR-520 toll bridge) showed a consistent drop in traffic volume in March, followed by a gradual increase in April. Figure 10 presents an example of the daily traffic volume trends on I-5 in downtown Seattle.

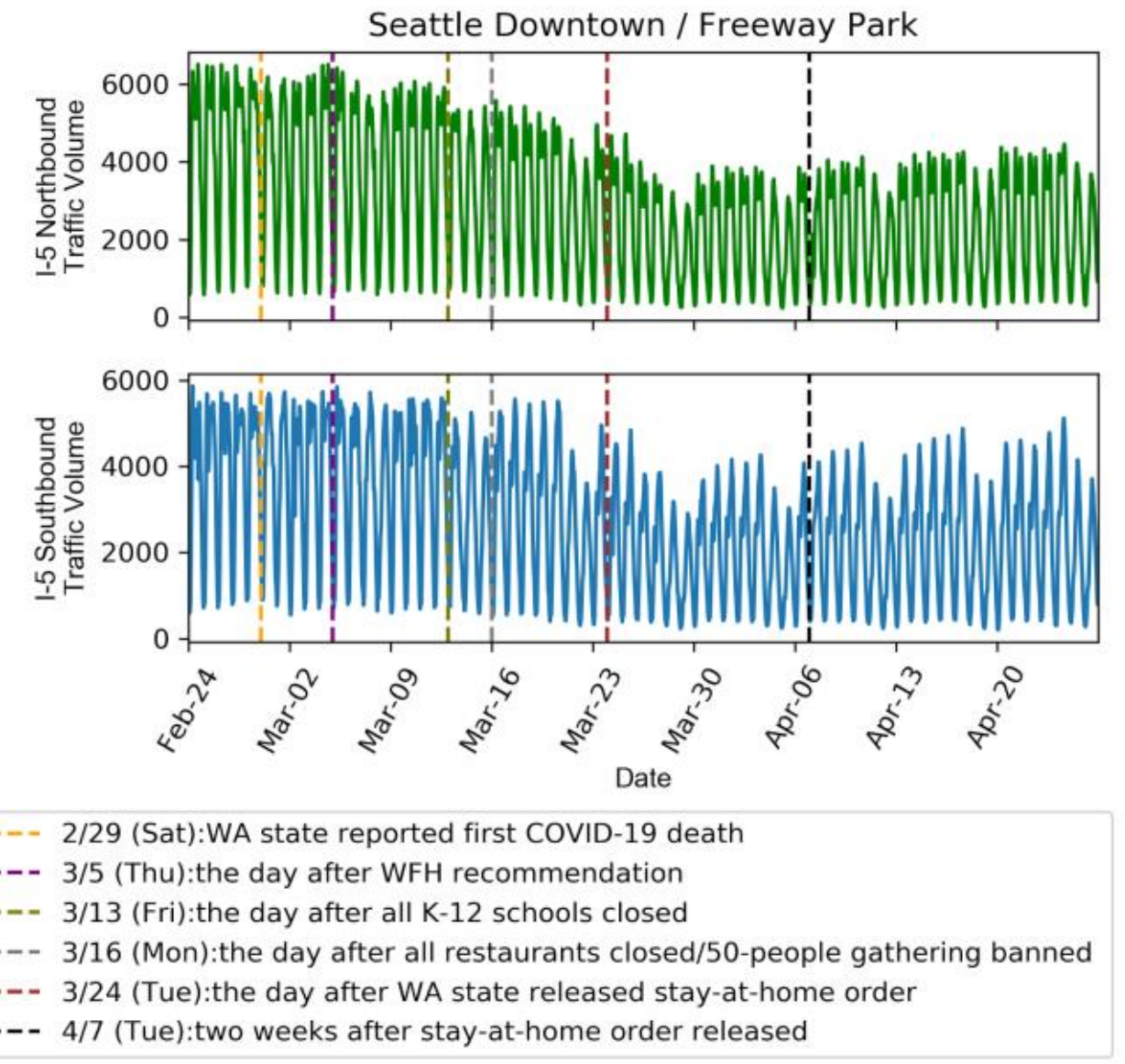

**Figure 10** Traffic volume on I-5 (downtown Seattle)

Using the week of March 30 to April 5 as a baseline (as it observed the lowest traffic volume), traffic volume at the I-5 Downtown Seattle station recorded a 6.6%, 11.9%, and 14.8% increase in the weeks of April 6, 13, and 20, respectively. Similar trends have been observed for the other two stations. The continuous increasing trend of traffic volume indicates that the impact of traffic in Seattle is slowly returning to normal levels despite stay-at-home restrictions remaining in place.

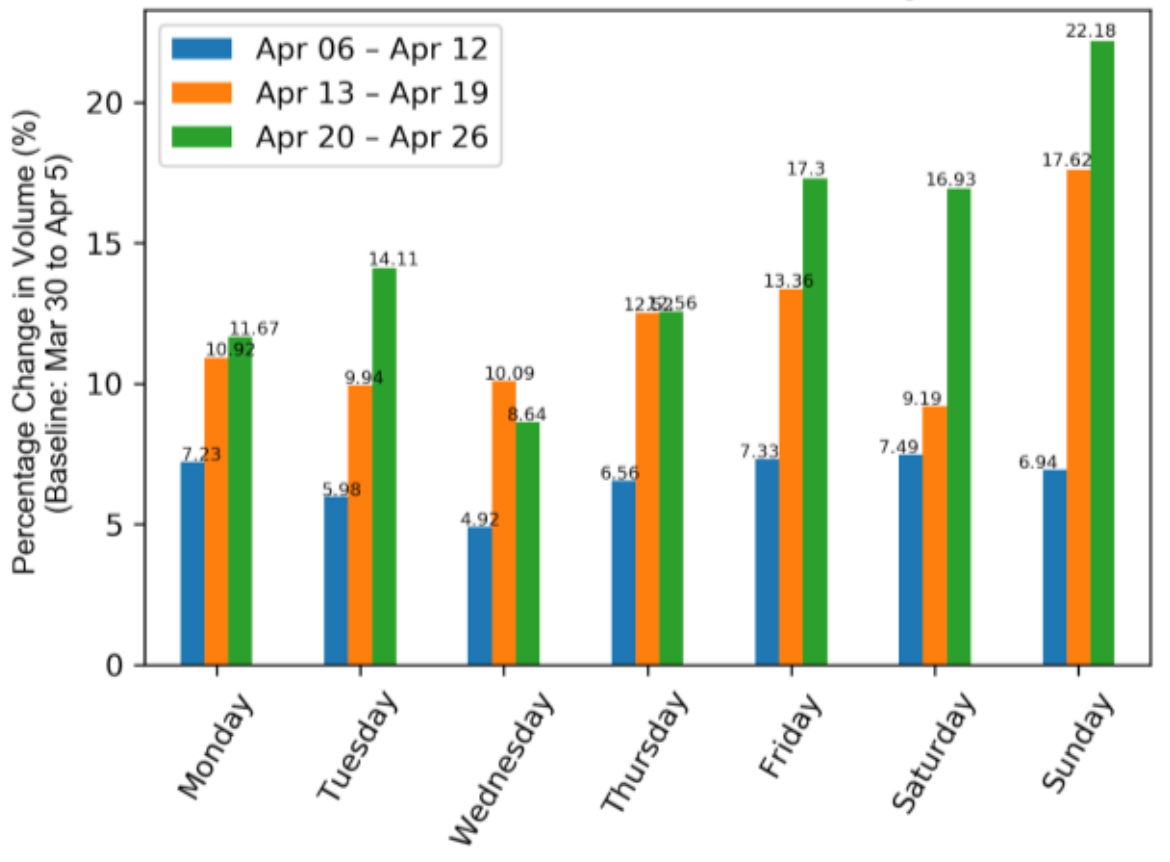

**Figure 11** Traffic volume change on I-5 (downtown Seattle)

## Transit Ridership

According to Transit app data, public transit demand of Seattle began to decline on February 29, 2020, the same day when Washington state reported the first COVID-19 death. The demand continuously decreased until March 26 (a 79% drop compared to a normal day), and demand has remained low since. The increase in traffic volumes while public transit ridership remains low indicates a potential shift in mode choice attitudes and preference for non-public modes of travel.

## Bicycling Volume

Bicycling, already an important mode of travel in Seattle, is increasing. Figure 12 shows the total bike counts near Fremont Bridge between mid-February and late-April. From mid-February to early-March, higher counts (+3,500) are observed on weekdays while fewer counts (~1,000) on weekends. During the first two weeks of March, the number of bikes gradually decreased, corresponding to work-from-home recommendation made on March 5. Counts reached the lowest level at the end of March, which is consistent with the pattern change in travel time and traffic volume. In April, bike counts grew to nearly 2,100, without a clear weekday/weekend pattern, possibly due to an increase in the use of bikes for recreational purposes and as a result of the "Stay Healthy Streets closures" program.

Seattle opened 6 miles of Stay Healthy Streets on April 24, 2020, creating a total of 9 miles to allow for safe social distancing building off of the 196 miles of Seattle's bike facilities and trails. Initial observations indicate bike travel was up 299% from 2017[iii] along Stay Healthy Streets in the Central District, which may contributed to the observed bike trend in Figure 12. SDOT is aiming to convert approximately 15 miles to Stay Healthy Streets in the coming weeks.

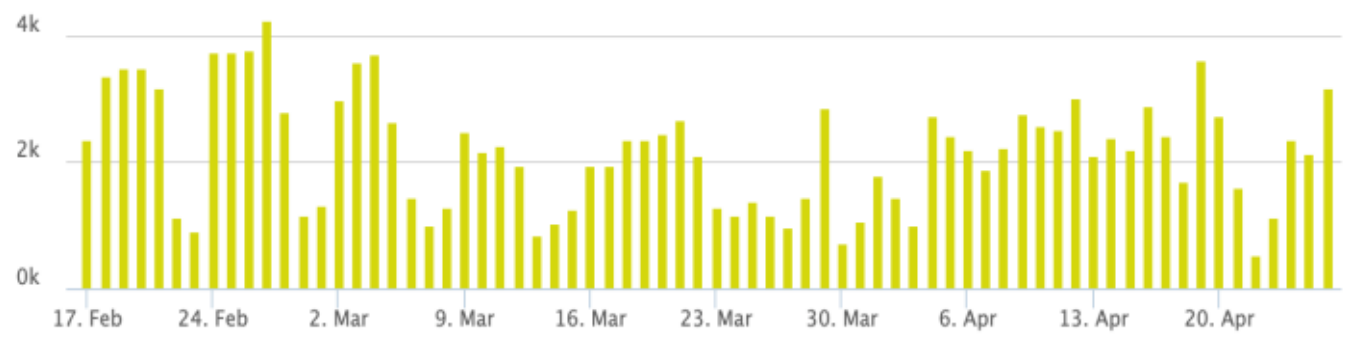

**Figure 12** Seattle bike counts near Fremont Bridge (Figure source: Seattle DOT bike counters)

## A Glance at Multiple Cities in China

As a way to better understand what recovery might look like, subway ridership and mode shift in multiple cities in China were investigated. As shown in Figure 13, subway ridership in major cities in China has been gradually increasing as cities reopen, but still remain well below normal. Most cities in China reopened around February 10, except Wuhan, the Chinese epicenter of the COVID-19 outbreak, which reopened its subway system in three phases from March 28-April 22.

In Shanghai and Guangzhou, an average rate of 40% and 63% of ridership (compared to the days before the outbreak) were restored after 1-month and 2-month of reopening, respectively. Beijing experienced a stricter strategy on reopening including a mandate to maintain subway car occupancy below 50% of its maximum capacity. This led to a slower recovery with a 20% and 40% restoration of subway ridership one and two months after reopening. Similarly, Wuhan has only restored 17% of its typical ridership one month after reopening. Other social distancing enforcement strategies such as limiting the amount of people entering subway stations during peak hours, are likely also contributing to the slow recovery.

A mode shift towards walking and cycling was also observed after the city reopened. According to a survey conducted by the Institute for Transportation and Development Policy (ITDP) in early March in Guangzhou, 40% of previous transit commuters had shifted to private cars, taxis, and ride-hailing, with more to walking and biking. Beijing's bike share systems saw an increase in usage by 150% from February 10 to March 4, 2020, one month after reopening. Continued monitoring of the mode shift and repositioning of transit services with new health standards can provide insight during the anticipated upcoming recovery period in the US.

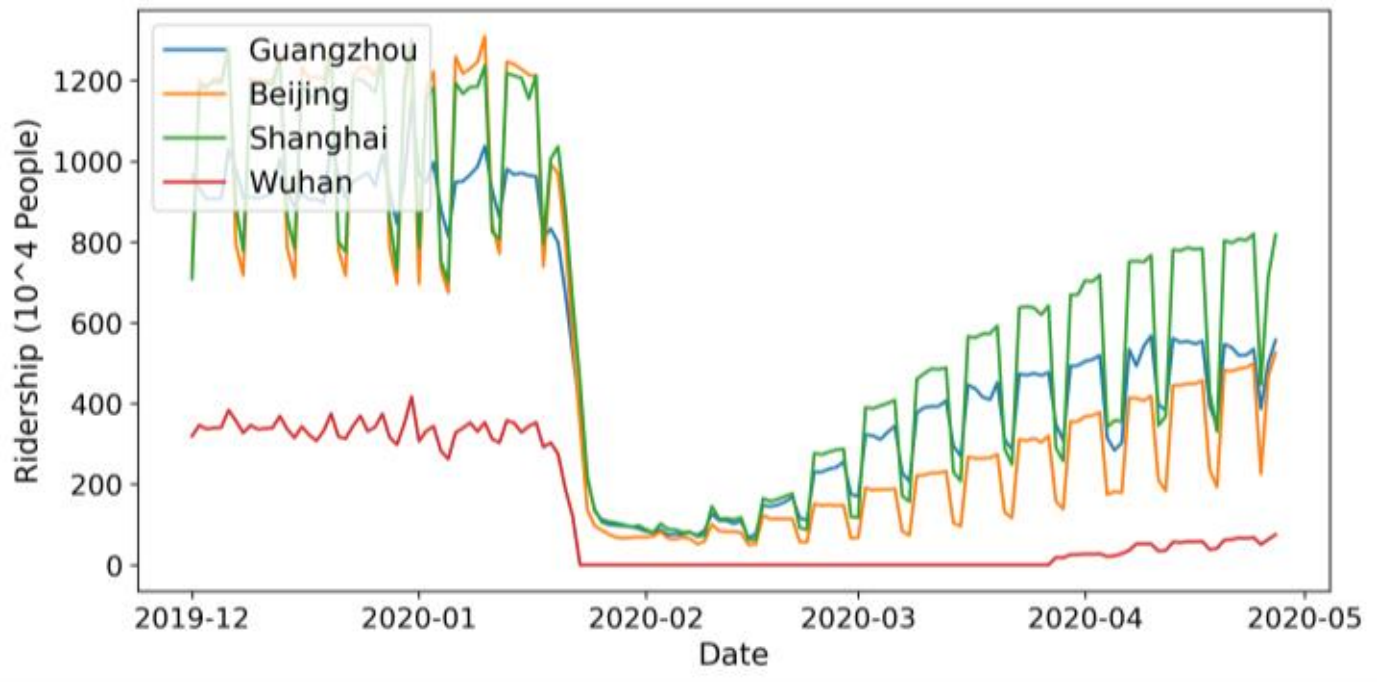

**Figure 13** Subway ridership changes in Guangzhou, Beijing, Shanghai and Wuhan, China

## Summary of Findings

NYC is still grappling with the COVID-19 pandemic but a reopening of the city may soon occur. Data from Seattle, which is potentially further along in the recovery shows that while vehicular traffic volume and bicycling are starting to increase, transit usage remains low. Moreover, transit ridership and mode shift in multiple cities in China, several months ahead in recovery, show that the recovery of transit ridership will be slow.

March marked the first time protracted period of congestion disappeared across many U.S roadways. The data now shows what "light" traffic condition looks like for the typical congested cities and can be used as a reference baseline for future studies of the resilience of the system.

This paper reflects the Center's perspective as of May 3, 2020 based on data collected in April 2020. As the world continues to adjust to the new reality of COVID-19, C2SMART researchers are continuing to collect data, including perishable mobility, safety, and behavior data, and will continue to monitor these trends and regularly update findings as some states are moving to reopen.

## Data Sources

- NYC Department of Transportation (DOT), Real-Time Traffic Speed Data and Traffic Speeds Map
- INRIX, Traffic Speed Data
- Metropolitan Transportation Authority (MTA), Turnstile Data and hourly traffic on MTA bridges and tunnels
- Citi Bike, Citi Bike System Data
- New York Police Department (NYPD), Motor Vehicle Collisions - Crashes
- Google, travel time information (extracted from Google Directions API)
- Washington State DOT, Traffic volume counts
- Transit App, Subway ridership demand
- Seattle Department of Transportation (SDOT), Street closure information and bike counts
- Shanghai, Beijing, Guangzhou and Wuhan Metro official Weibo account, Metro Ridership
- Institute for Transportation and Development Policy, Travel mode choice survey in Guangzhou, China

*Note: all data is preliminary and subject to change.*

*For more information please contact* c2smart@nyu.edu

*For more data visualization, please visit our COVID-19 Dashboard:* c2smart.engineering.nyu.edu/covid-19-dashboard/

[i] The Wall Street Journal, The Roads Are Quieter Due to Coronavirus, but There Are More Fatal Car Crashes

[ii] Subway ridership data is for 3/02- 4/17/2020.

[iii] SDOT, Stay Healthy Streets continue this week and will add 3 new neighborhoods this weekend, https://sdotblog.seattle.gov/2020/04/23/stay-healthy-streets-continue-this-week-and-will-add-3-new-neighborhoods-this-weekend/